\newcommand{\vct}[1]{{\bf #1}}
\newcommand{\eq}[1]{Eq.~(\ref{eq:#1})}
\newcommand{\fig}[1]{Fig.~\ref{Fig:#1}}

\newcommand{\rv}{\ensuremath{\vct r}}
\newcommand{\Rv}{\ensuremath{\vct R}}

\newcommand\smaller[2][0.78]{{\scalefont{#1}#2}}

\documentclass[%
preprint,
superscriptaddress,
showpacs,
nopreprintnumbers,
nobibnotes,
 amsmath,amssymb,
 aps,
pra,
]{revtex4-1}

\usepackage{graphicx}
\usepackage{dcolumn}
\usepackage{bm}
\usepackage[usenames,dvipsnames]{color}
\usepackage{mathrsfs}
\usepackage{scalefnt}

\begin{document}
\let\xxxhat\hat
\let\xxxvec\vec
\renewcommand{\hat}[1]{{\boldsymbol {\xxxhat {#1}} }}
\renewcommand{\vec}[1]{\boldsymbol {#1}}


\title{Contact interaction in an unitary ultracold Fermi gas}

\author{Renato Pessoa}
\affiliation{Instituto de F\'{\i}sica,
Universidade Federal de Goi\'as - UFG,
74001-970 Goi\^ania, GO, Brazil} 
\affiliation{Department of Physics,
Arizona State University, Tempe, AZ, 85287 USA}
\author{S. Gandolfi}
\affiliation{%
Theoretical Division, Los Alamos National Laboratory, Los Alamos, New
Mexico 87545, USA}
\author{S. A. Vitiello}
\affiliation{Instituto de F\'{\i}sica Gleb Wataghin,
Universidade Estadual de Campinas - UNICAMP,
13083-970 Campinas, SP, Brazil}
\affiliation{%
JILA, National Institute of Standards and Technology and Department of
Physics,\\
University of Colorado, Boulder, Colorado 80309-0440, USA}
\author{K. E. Schmidt}
\affiliation{Department of Physics,
Arizona State University,
Tempe, AZ, 85287 USA}
\affiliation{Instituto de F\'{\i}sica Gleb Wataghin,
Universidade Estadual de Campinas - UNICAMP,
13083-970 Campinas, SP, Brazil}

\date{\today}

\begin{abstract}
An ultracold Fermi atomic gas at unitarity
presents universal properties that in the dilute limit
can be well described by a contact interaction.
By employing a guiding function with correct boundary conditions and making
simple modifications to the sampling procedure we are able to calculate
the properties of a true contact interaction with
the diffusion Monte Carlo method.
The results are obtained
with small variances. Our calculations for the Bertsch and
contact parameters 
are in excellent agreement with published experiments.
The possibility of using a more faithful description of ultracold atomic gases
can help uncover additional features of ultracold atomic gases.
In addition, this work paves the way to perform quantum Monte Carlo calculations
for other
systems interacting with contact interactions, where the description
using potentials with finite effective range might not be accurate.
\end{abstract}

\maketitle


\section{\label{sec:intro}Introduction}

Systems formed by fermions have
many-body  properties that
are of central
importance for understanding observed phenomena in many fields of physics.
These fields include
ultracold gases, condensed matter, and nuclear physics.  The
possibility of handling
ultracold atomic Fermi gases, in a very precise way,
has allowed testing quantum many-body theories
in an unprecedented set of experimental conditions.

Ultracold Fermi gases can be tuned
from weakly interacting to strongly correlated regime
by application of magnetic fields
near a Feshbach
resonance~\cite{chi10}.
When the interaction has diverging scattering length, the unitary
limit, the system presents universal properties, \textsl{i.e.}, it does not
depend on the nature of the interactions. The system universality
allows one to study
the crossover from the Bardeen-Cooper-Schrieffer (BCS)
superfluid state
to the Bose-Einstein condensed (BEC) state, in general~\cite{gio08}.

Countless efforts were made and continue to be made
\cite{wei15,*bar14,*sid13} to uncover the many aspects involved in the
observed phenomena presented by the ultracold Fermi
gases.  In the present work we investigate the unitary limit of this
system at
the crossover from BCS to the BEC regime with an $s$-wave contact
interaction.


Interactions of two neutral atoms are not always easy to describe
accurately.
However in the dilute regime,
interactions
can be well represented by two-body collisions
using a contact potential.
Nevertheless a straightforward consideration of this type of potential
makes theoretical investigations problematic because when two particles
approach one another the wave function diverges.
This difficulty is usually avoided by adopting pseudopotentials of the
P\"oschl-Teller, hard sphere, square well, or other forms~\cite{for12}.  In this fashion,
valuable insights have come from quantum Monte Carlo methods
\cite{car03,ast04,lob06,gan11}, despite the fact that
finite-range potentials lead to incorrect scattering properties, which are
fundamental quantities of these systems. The resulting calculations
must therefore include an additional extrapolation to zero range. Since
the trial wave functions diverge in this limit, the extrapolations
are not well behaved in this limit.

It is not just a matter of principle or of interest in itself to avoid
using finite-range pseudopotentials
to describe the two-body interaction of
ultracold Fermi gases.  For instance,
it is important to avoid the possible influence of
the true ground state of the P\"oschl-Teller model system,
since it may have tightly-bound
states highly dependent on the chosen range. For
repulsive interactions, there are still open questions about the ferromagnetic
character of the ground state and what kind of ferromagnetic transition
the system undergoes in this case \cite{bug14,cha11,pil10}. The possibility of
simulating Fermi atomic gases considering a contact interaction will help
solve questions like those previously mentioned.
On the other hand, studies of Bose systems, including Bose-Fermi
mixtures, have been done only using finite range interactions in
quantum Monte Carlo calculations, see for example Refs.~\cite{ros14,ber13},
introducing possible bias in the calculation.

The contact interaction as we have considered
allows the quantities of interest to be
obtained without the additional
burden of performing
extrapolations to zero-range interactions.
This is useful in a twofold way.
It can help understand how
previous results
might have been affected by the use of finite range potentials,
and also because the calculations become simpler.
Moreover,
the results we present depend on relatively small changes of the standard
diffusion Monte Carlo (DMC)
algorithm.
Additionally, we show how to compute the two-body propagator for
particles interacting through a contact potential,
which is an interesting result in itself.

\section{Methodology}

The system we study consists of $N$ fermionic particles described through the Hamiltonian
\begin{equation}
H = -\frac{\hbar^2}{2m}\left[ \sum_{i}^{N/2}\nabla_{i}^2 + \sum_{i'}^{N'/2}\nabla_{i'}^2 \right] + \sum_{i,i'} v(r_{i,i'}) \label{eq:Hamiltonian}
\end{equation}
where the first terms are the kinetic energies of the up-spin (unprimed index) and down-spin (primed index) particles and the last term is the zero-range interatomic potential. Here we focus on the unpolarized system, and $N/2$ particles are spin-up, and $N/2$ are 
spin-down. The simplest way the solve \eq{Hamiltonian} is to
introduce a trial variational wave function $\Psi_T(R)$, where $R\equiv
\{\mathbf{r}_1,\mathbf{r}_1',\cdots, \mathbf{r}_{N/2},\mathbf{r}_{N'/2}
\}$, and minimizing the expectation value of $H$~\cite{fou01}.  Typically,
one samples $M$ configurations from the probability density proportional to
$\left |\Psi_T\right |^2$
and average the local energy. The value of the variational Monte Carlo
energy $E_{VMC}$ is a ground state upper bound and it  normally depends
on the quality of the trial wave function. Beyond the VMC calculation
using the diffusion Monte Carlo(DMC) one can project out the
lowest state of the system from $\Psi_T$.


The Schr\"odinger equation can be written as a diffusion equation in imaginary time $\tau=it/\hbar$,
\begin{equation}
-\frac{\partial\psi(R;\tau)}{\partial \tau} = \left( H - E_T\right)\psi(R;\tau)
\end{equation}	
where $E_T$ is introduced to stabilize the norm of $\psi(R;\tau)$ in the limit of $\tau\rightarrow\infty$. It is convenient to expand the trial wave function in the eigenstates $\{\psi_i\}$ of $H$ and make $E_T\approx E_0$. In that case, it is straightforward to show~\cite{cha04,pes15} that we project out the $\psi_0$ evolving $\Psi_T$ in the imaginary time. In practice, the total Monte Carlo time is divided into $n$ small time steps such that $\tau=\Delta\tau \cdot n$ and, the exact wave function is propagated like
\begin{equation}
\psi(R;\tau) =\lim_{n\rightarrow\infty} \left[ e^{-\left( H - E_0\right)\Delta\tau}\right]^n \Psi_T(R)\;\;.
\end{equation}
In the DMC method the projected state at large imaginary time is the
lowest energy state
not orthogonal to the trial function.


The zero-range interaction in $s$-wave is enforced by using an
importance function that satisfies the Bethe-Peierls condition 
$\left({1}/{r_{i,j'}} - {1}/{a}\right)$ when $r_{i,j'}\rightarrow 0$,
where 
$a$ is the two-body scattering length.
At unitarity,
($a \rightarrow \infty$)
the Bethe-Peierls condition reduces simply to the wave funtion being
proportional to
the inverse modulus of the pair relative distance at small separations.
This approach allows us to treat the system with the zero-range pseudopotential
as formed from
pairs of free particles subject
to the correct boundary conditions when
a pair separation distance goes to zero.

The importance function we use can be written as:
\begin{eqnarray}
\label{eq:bcs_wave}
\Psi_T = \prod_{ij'}f(r_{ij'})\Phi_{BCS}\;\; ; \\
\Phi_{BCS} = \mathcal{A} \left[ \phi(\mathbf{r}_{11'})\phi(\mathbf{r}_{22'})\cdots \phi(\mathbf{r}_{N/2\;N'/2}) \right]
\end{eqnarray}
where the Jastrow pair function $f(r)$ correlates the unlike spin pairs
of the system. The pair function is chosen to satisfy the Bethe-Peierls
condition, and the
antisymmetric BCS function $\Phi_{BCS}$ is well behaved at small pair
separations and defines the nodal surface
structure~\cite{car03}. The operator $\mathcal{A}$ antisymmetrizes like
spin pairs.
Here
we use the general form for the BCS part where the
pairing functions are written like a set of plane waves
\begin{equation}
\label{eq:pair}
\phi(r) = \sum_{j=1}^{N_S} c_j e^{i\vct k_j\cdot \vct r} \,,
\end{equation}
and $c_j$ are variational parameters. The $c_j$ coefficients with
the same magnitude $\vec k_j$ are equal.

These orbitals have the same form of previous works~\cite{pes15}.
The short-range pairing function of
Refs.~\onlinecite{car03,gan11}
is not included since the boundary condition from the potential
is enforced by the Jastrow factor. We have considered $N_S=20$ shells in the
guiding function and have obtained converged energies for both
variational and DMC calculations.
The coefficients entering in the pairing orbitals have been optimized 
as described in~\cite{gan11} using the stochastic reconfiguration method 
\cite{sor01}. The function $\Phi_{BCS}$
is projected on the subspace with fixed number of particles $N$
\begin{equation}
|BCS\rangle
=\prod_i (u_i + v_i a^\dagger_{\vct k_i\alpha} a^\dagger_{-\vct k_i\beta})
|0\rangle
\end{equation}
If for $k_i>k_F$ all the $v_i$ are equal to zero this function reduces to
a product of two
Slater determinants of plane waves\cite{pes15}.
We call this the
Jastrow-Slater
wave function; it was
used in our previous work~\cite{pes15}.

The Jastrow pair function $f(r)$ in \eq{bcs_wave} correlates the unlike spin
pairs, and we take
\begin{equation}
f(r)=\frac{d \cosh(\lambda r)}{r \cosh(\lambda  d)}
\end{equation}
with $f(r\ge d) = 1$,
the parameter $\lambda$ is chosen to make $f$ and its first
derivative continuous at
the healing distance $r=d$.
Its value
is of the order of the inverse
interparticle distance and
is determined with a variational calculation.
It is important that the Jastrow pair function has the correct boundary condition at short distances.


Our variational calculations are performed as in Ref.~\onlinecite{pes15}. The
variance of the energy of this trial function with the usual VMC
method is not well behaved. This can be seen by looking at the form
of the trial function when a pair separation is small. For example, if
up-spin particle $1$ and down-spin particle $1'$ are close together,
the Jastrow factor $f(r_{11'})$ goes like $r_{11'}^{-1}$. The term in the
Hamiltonian where $-\frac{\hbar^2}{2m}(\nabla_1^2+\nabla_{1'}^2)$ operates
on this gives zero except at the origin where it cancels the contact
interaction, so that part of the local energy is well behaved. The problem
terms are those like
$\vec \nabla_1 f(r_{11'})\cdot \vec \nabla_1 \frac{\Psi_T(R)}{f(r_{11'})}$.
This term goes like
$\frac{\vec r_{11'}}{r_{11'}^3} \cdot\vec\nabla_1 \frac{\Psi_T(R)}{f(r_{11'})}$
at small distances. The $r_{11'}^2$ term in the volume element as well
as the angular integration makes this term give a well behaved contribution
to the energy as the pair separation goes to zero, however, this
term is squared in the energy variance, and the variance diverges.

The exact ground state wave function must, of course, have zero variance
with $\frac{H\Psi_0(R)}{\Psi_0(R)} = E_0$ independent of $R$. This means
that the exact wave function must have additional terms which cancel these
divergences. These would have the form of three-body correlations which would
cancel the divergences from terms like
$\vec \nabla_1 f(r_{11'}) \cdot \vec \nabla_{1} f(r_{1j'})$, and backflow
terms to cancel divergences from terms like
$\vec \nabla_1 f(r_{11'}) \cdot \vec \nabla_1 \phi(r_{1j'})$.
Such terms would have to be constructed in such a way as to not spoil
the necessary boundary conditions. A hierarchy of such terms may be required
to obtain a well behaved variance of the local energy.

Since the integrated energy is well behaved, we have chosen to attack the
problem by modifying the sampling to control the variance. The key insight
is that for $r_{11'} \rightarrow 0$, interchanging the positions of particles
$1$ and $1'$ reverses the sign of the gradient $\vec \nabla_1 f(r_{11'})$,
but does not change the rest of the trial wave function. We therefore
modify the standard Metropolis algorithm to include moves which interchange
the positions of the closest pair of unlike spin particles. If the
pair remains the closest pair after interchange, we accept this move
with the heat bath probability for interchange
\begin{equation}
P_{int} = \frac{\Psi_T^2(R_{int})}{\Psi_T^2(R)+\Psi_T^2(R_{int})}
\end{equation}
where $R_{int}$ are the coordinates with the closest pair interchanged.
We use the method of expected values to evaluate the energy, after such
a trial move, so that the energy contribution is
$E_L(R_{int}) P_{int} + E_L(R)(1-P_{int})$. In the limit of small
pair separations, $P_{int} \rightarrow \frac{1}{2}$, and the diverging
contributions cancel.

The diffusion Monte Carlo calculations have the same diverging terms in
the local energy, and we employ a similar technique to control the variance.

The propagation equation in imaginary time including $\Psi_T$
as an importance function
\begin{widetext}
\begin{equation}
\Psi_T(R)\psi(R;\tau+\Delta \tau) = \int d^3 R'\frac{\Psi_T(R)}{\Psi_T(R')}G(R,R';\Delta \tau) \Psi_T(R')\psi(R',\tau). \label{eq:psit}
\end{equation}
\end{widetext}
Since the zero-range interatomic potential is a delta function, the usual Trotter-Suzuki decomposition of the propagator $e^{-H\Delta
\tau}$ is not adequate. The walkers are instead sampled from $\frac{\Psi_T(R)}{\Psi_T(R')}G(R,R';\Delta \tau)$~\cite{pes15} and we essentially have only to deal with the kinetic energy term
of the Hamiltonian. 
The short time propagator we use is evaluated using the pair
product form from the 
two-body propagators $g$
\begin{equation}
\label{eq:gr}
G(R',R) =
G_0(R',R) \prod_{i<j}
\frac{g(\rv'_{i},\rv'_{j};\rv_{i},\rv_{j})}
{g_0(\rv'_{i},\rv'_{j};\rv_{i},\rv_{j})} \,,
\end{equation}
where
$G_0$ and $g_0$
are the free particle and
the free pair propagators, respectively.
Note that for pairs with the same spin $g=g_0$.
For pairs with opposite spin
$g$ can further be written as
$g_{\rm rel} \times g_{\rm cm}$,
the product of
the relative times
the center of
mass propagators of the pair
\begin{equation}
\label{eq.gpair}
g(\rv'_{i},\rv'_{j'};\rv_{i},\rv_{j'})
=g_{\rm rel}(\rv'_{ij'};\rv_{ij'})
g_{\rm cm}(\Rv'_{ij'};\Rv_{ij'}),
\end{equation} 
where
$\Rv_{ij'} = (\rv_{i} + \rv_{j'})/2$ is the center of mass of the pair.
In our approach it is necessary to write the full propagator as above.

The centers of mass propagate like free particles~\cite{pes15}. On
the other hand, the two-body propagator is a Green's function that can
be constructed from the normalized solution of the of the scattered
\textit{s}-wave function as employed in other papers~\cite{pes15,yan15}
\begin{equation}
g_{\rm rel}(\mathbf{r},\mathbf{r}';\Delta\tau) = \sum_n \varphi_n(r) e^{-\frac{\hbar^2k_n^2}{m}\Delta\tau}\varphi_n^{*}(r') \label{eq:Grr}
\end{equation}
where $\{ \varphi \}$ is the complete set of eigenstates of the two-body
Hamiltonian. Since the interaction is only in the s-wave, we can
separate into partial waves, and the s-wave
contribution for scattering length $a$ becomes
\begin{eqnarray}
g_s(r,r',a) &=& \frac{1}{4\pi^2 r r'} {\rm Re}
\int_0^\infty dk \left [
-\frac{(1-ika)^2}{1+k^2a^2}e^{ik(r+r')}+e^{ik(r-r')} \right]
e^{-\frac{\hbar^2 k^2 \Delta t}{m}} +{\rm \ bound\ state}\,,
\nonumber\\
\end{eqnarray}
where for positive $a$, the bound state contribution should be included.
The integrals can be done straightforwardly in terms of gaussians
and error functions,
\begin{eqnarray}
g_s(r,r',0) &=& \frac{1}{8\pi^2 r r'} \sqrt{\frac{m \pi}{\hbar^2 \Delta t}}
\left [ -e^{-\frac{m}{4\hbar^2 \Delta t} (r+r')^2}
+e^{-\frac{m}{4\hbar^2 \Delta t} (r-r')^2} \right ]
\nonumber\\
g_s(r,r',-\infty) &=&
\frac{1}{8\pi^2 r r'} \sqrt{\frac{m \pi}{\hbar^2 \Delta t}}
\left [e^{-\frac{m}{4\hbar^2 \Delta t} (r+r')^2}
+e^{-\frac{m}{4\hbar^2 \Delta t} (r-r')^2} \right ]
\nonumber\\
g_s(r,r',a) &=& g_s(r,r',-\infty)
-\frac{1}{4\pi r r' |a|}e^{\frac{\hbar^2\Delta t}{m a^2}-\frac{r+r'}{a}}
{\rm erfc}\left (
\sqrt{\frac{\hbar^2\Delta t}{m a^2}}
-\frac{r+r'}{2a} \sqrt{\frac{ma^2}{\hbar^2\Delta t}}
\right)
\end{eqnarray}
where any bound state contribution needs to be added.
Here we are primarily interested in the unitary case $a=-\infty$ where
the relative coordinates propagator is particularly simple~\cite{car12}
\begin{equation}
g_{\rm rel}(\mathbf{r},\mathbf{r}';\Delta\tau) = g_{\rm rel}^0(\mathbf{r},\mathbf{r}';\Delta\tau) + \frac{1}{4\pi^2 rr'} \sqrt{\frac{m\pi}{\hbar^2\Delta\tau}}e^{-\frac{m}{4\hbar^2\Delta\tau}\left(r + r' \right)^2}
\label{eq:grel}
\end{equation}
where the first term is a free-particle propagator for the relative
distances and the last one is the contribution of the \textit{s}-wave
scattering.

The sampling of the importance sampled
propagator $\frac{\Psi_T(R')G(R',R)}{\Psi_T(R)}$
is accomplished by approximating
it, sampling the approximation, and using the ratio of
$\frac{\Psi_T(R')G(R',R)}{\Psi_T(R)}$ to the approximation as a weight.

We first construct what we call the independent pair
propagator $G_{\rm ip}(R',R)$.
We sort the unlike spin pair distances, and first select the closest pair.
We then eliminate all pairs that contain the closest pair's particles.
We repeat this process on the remaining pairs. The result is a list of
independent pairs. The independent pair propagator is the
product of the pair propagators Eq. \ref{eq.gpair} for these independent
pairs. From the form of the relative pair propagator, we see that
if the initial separation is much larger than
$\sqrt{\frac{4\hbar^2\Delta \tau}{m}}$, the propagator becomes the
free particle propagator. Furthermore for large separations, the divergences
in the trial function can be neglected. Therefore we introduce a cutoff
parameter, so that if the separation is larger than this parameter, we
sample the pair from the free particle propagator. If it is less than this
cutoff, we approximate the trial function by the Jastrow factor for that
pair, and for these separations, we take its asymptotic value,
given by the Bethe-Peierls condition.
We sample the center of mass part of the pair propagator from
the noninteracting center of mass gaussian, and the relative separation from
$\frac{r_{ij'}}{r'_{ij'}} g_{\rm rel}(\vec r_{ij'}',\vec r_{ij'})$.
The details of this sampling are given in the appendix. This general method can
be readily extended to scattering lengths away from unitarity.

The value of the cutoff parameter does not affect our results, and for
reasonable values has very little effect on the variance.
We
denote the sampled configuration $R_1 = R + \Delta R$ obtained as
described above by
$\mathscr{P}_{\rm ip}(R_1,R) =
G_{\rm ip}(R,R')
\prod_{i<j}
\frac{r_{ij'}}{r'_{ij'}}$,
where the product of the Bethe-Peirls condition is only over the pairs
that are within the cutoff distance.

To include importance sampling, we use the antithetic ``plus-minus'' sampling
often used in nuclear quantum Monte Carlo calculations\cite{car15}, for the
center of mass variables, and the relative coordinates beyond the cutoff.
For these, the gaussians have the same probability of taking the opposite
sign. Therefore, it is equally probable for us to have sampled the
configuration $R_2 = R - (R_1 - R)$.

A divergence in the local energy at the sampled point $R_1$ can
occur exactly as in the variational calculations. To avoid this,
two additional
configuration are also considered. These are obtained by
interchanging the closest pair from $R_1$ and $R_2$.
Finally, a new configuration is chosen
among the $R_i$ according to
\begin{equation}
\frac{\frac{\Psi_T(R')}{\Psi_T(R)}G(R',R)}
{\sum_j  \frac{\Psi_T(R_j)}{\Psi_T(R)}G(R_j,R)}
\sum_j \mathscr{P}_{\rm ip}(R_j,R).
\end{equation}
By performing this choice, the importance sampled
$\frac{\Psi_T(R')}{\Psi_T(R)}G(R',R)$ is recovered by
through the weight
\begin{equation}
\label{eq:we}
\mathscr{W}(R') = \frac
{\sum_j  \frac{\Psi_T(R_j)}{\Psi_T(R)}G(R_j,R)}
{\sum_j \mathscr{P_{\rm ip}}(R_j,R)}.
\end{equation}


\section{Results and discussion}

In the unitary limit,
the resonant character of the interactions of
a Fermi gas makes the system have
only two possible energy scales: the chemical potential $\mu$ and the
Fermi energy $E_F$. Therefore these two quantities must be proportional,
$\mu=\xi E_F$.
As the temperature approaches zero,
the reduced chemical potential $\mu/E_F$
saturates
to the universal value $\xi$.
Of course, in this limit, $\mu$ converges to the system ground state
energy.
The value of $\xi$
has been accurately measured: $\xi = 0.376(4)$ \cite{ku12}.
However a more recent work  \cite{zur13}
suggests corrections to this value,
resulting in $\xi = 0.370(9)$.
If the atomic interaction is described by finite range pseudopotentials,
determining accurate values of $\xi$
requires a careful extrapolation to zero
effective range \cite{for12}. Our result for this quantity, also known as
the Bertsch parameter,
is $\xi = 0.390(2)$, obtained by simulating a system with 66 particles.
It is obtained in a straightforward manner, subject only to the
fixed-node approximation and finite size dependence. The
determined value is in reasonable agreement with the experimental one.
We have observed that there is only a small dependence of
this quantity on the number of particles in our
simulations, as also reported in Ref.~\cite{for11}.
The
energy we can obtain is in agreement with the best fixed-node diffusion Monte Carlo calculations performed using finite effective range interactions; in Ref.~\cite{car11} using the auxiliary-field
quantum Monte Carlo and a exact lattice technique, $\xi$ was determined as 0.372(5).

\begin{figure}
\begin{center}
\includegraphics[scale=0.44]{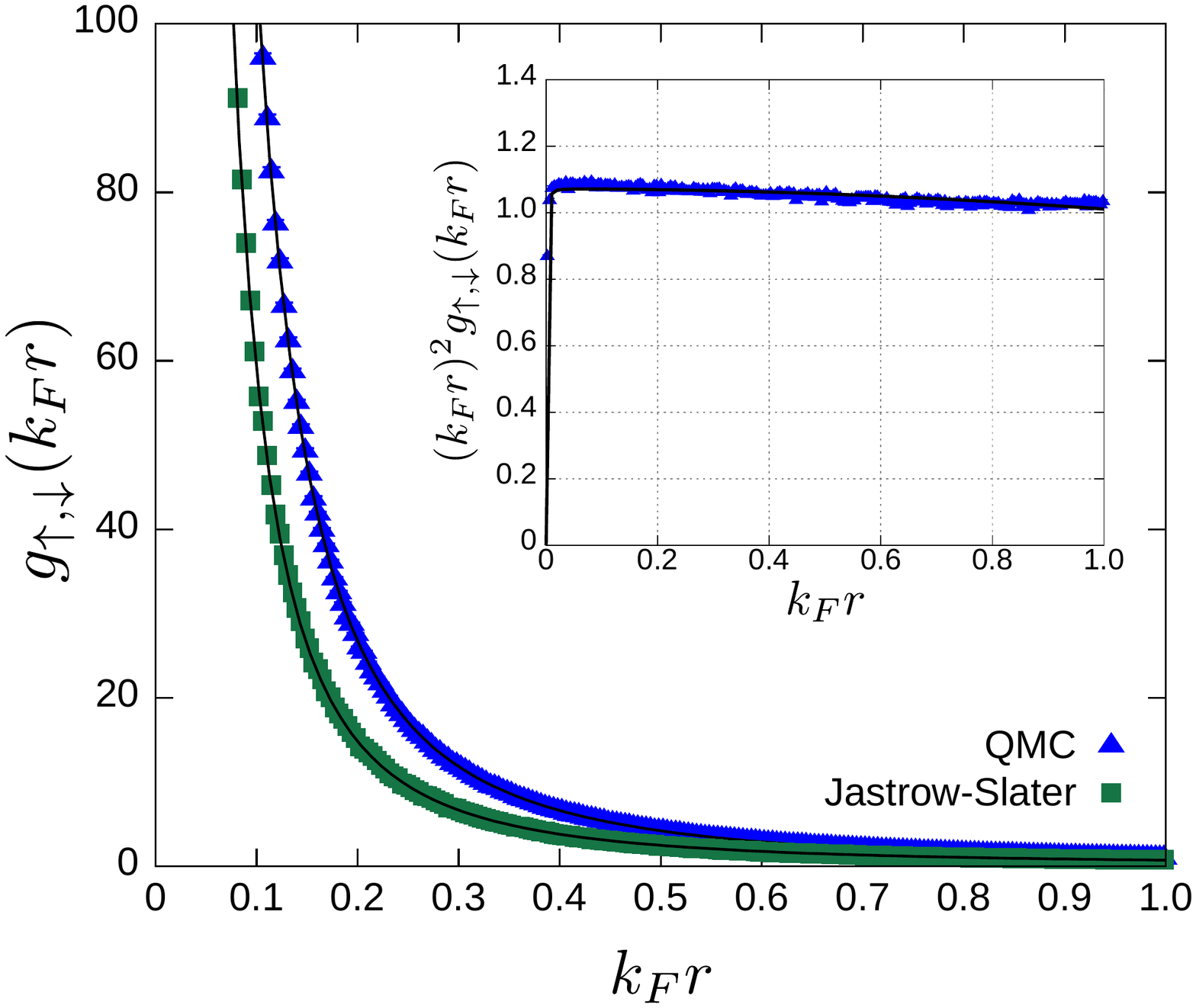}
\caption{\label{Fig:gr_un}(Color online) Pair distribution function of
unlike-spin pairs as a function of the distance.
Our quantum Monte Carlo (QMC) results are
for a system with 66 particles.
The solid line is the best fit of $b_0+b/(k_Fr)^2$
to the extrapolated points.
For completeness we have included
results for the Jastrow-Slater model
for a non superfluid system of particles
\cite{pes15}.
The inset presents the same distributions
multiplied by $(k_F r)^2$ as a function of the distance.}
\end{center}
\vspace{-5ex}
\end{figure}

The strong interacting Fermi gases described
by contact interactions obey a number of
universal relations characterized by a single parameter dubbed the
contact $C$ by
Tan~\cite{tan08a,*tan08b,*tan08c}. 
As shown by Zhang and Leggett \cite{zha09} the contact is able to enclose
all of the many-body physics.  The contact density
integrated in the whole space gives the contact, which is proportional to
the number of pairs with opposite spins that are close together. Its value can
also be computed straightforwardly from our calculations.  Before computing its
value it is useful to extract
a related constant $\zeta$
from the pair distribution function of unlike-spin
pairs as a function of the distance
presented in \fig{gr_un}.
At the unitary limit and for small distances 
\cite{gan11}
\begin{equation}
\label{eq:gab}
g_{\uparrow,\downarrow} (k_F r)
\rightarrow \frac{9\pi}{20} \frac{\zeta}{(k_f r)^2}.
\end{equation}
This is because the pair distribution function of particles with opposite spins
separated by small distances
satisfies in a first approximation
$g(r) \propto f^2(r)$.
%
The behavior of $g_{\uparrow,\downarrow}(k_F r)$ at
small distances
confirms with what we expect from \eq{gab} as we can verify
from the inset of \fig{gr_un}.
If we modify the fit by imposing $b_0=0$
we have estimated $\zeta = 0.755(1)$.
This value is slightly smaller that the one obtained with a fit where
$b_0$ is a free parameter. With this fit, it also
becomes more clear that a perfect
agreement between the DMC results and the fitted black line
in the inset of \fig{gr_un} occurs for small
values of $k_F r$.
The BCS
result is shifted to the right of the Jastrow-Slater, most probably
due to a large delocalization of the particles in the superfluid state.

The relation between the constant $\zeta$
and the contact parameter at unitarity is simple,
$C/k_F^4 = 2\zeta/5\pi$ \cite{gan11}.
However to make the comparison with experimental data easy, we report this
quantity in terms of the contact per unit volume given by
$\mathcal{C}/Nk_F = 3\pi^2 C /k_F^4$.
Our result, $\mathcal C$=2.848(1), is slightly below two recent measurements.
A Bragg spectroscopy experiment~\cite{hoi13} gives the value
3.06 $\pm$ 0.08 at the temperature $T/T_F = 0.08$.
A measurement
using radio-frequency spectroscopy gives 2.9 $\pm$ 0.3 at $(k_Fa)^{-1} = -0.08$
and $T/T_F = 0.18(2)$, a temperature slightly above the transition temperature 
$T_c$~\cite{sag15}.
Our computed value is closer to the experimental values than previous
results determined with a finite range potential~\cite{gan11}.

\begin{figure}
\begin{center}
\includegraphics[scale=.50]{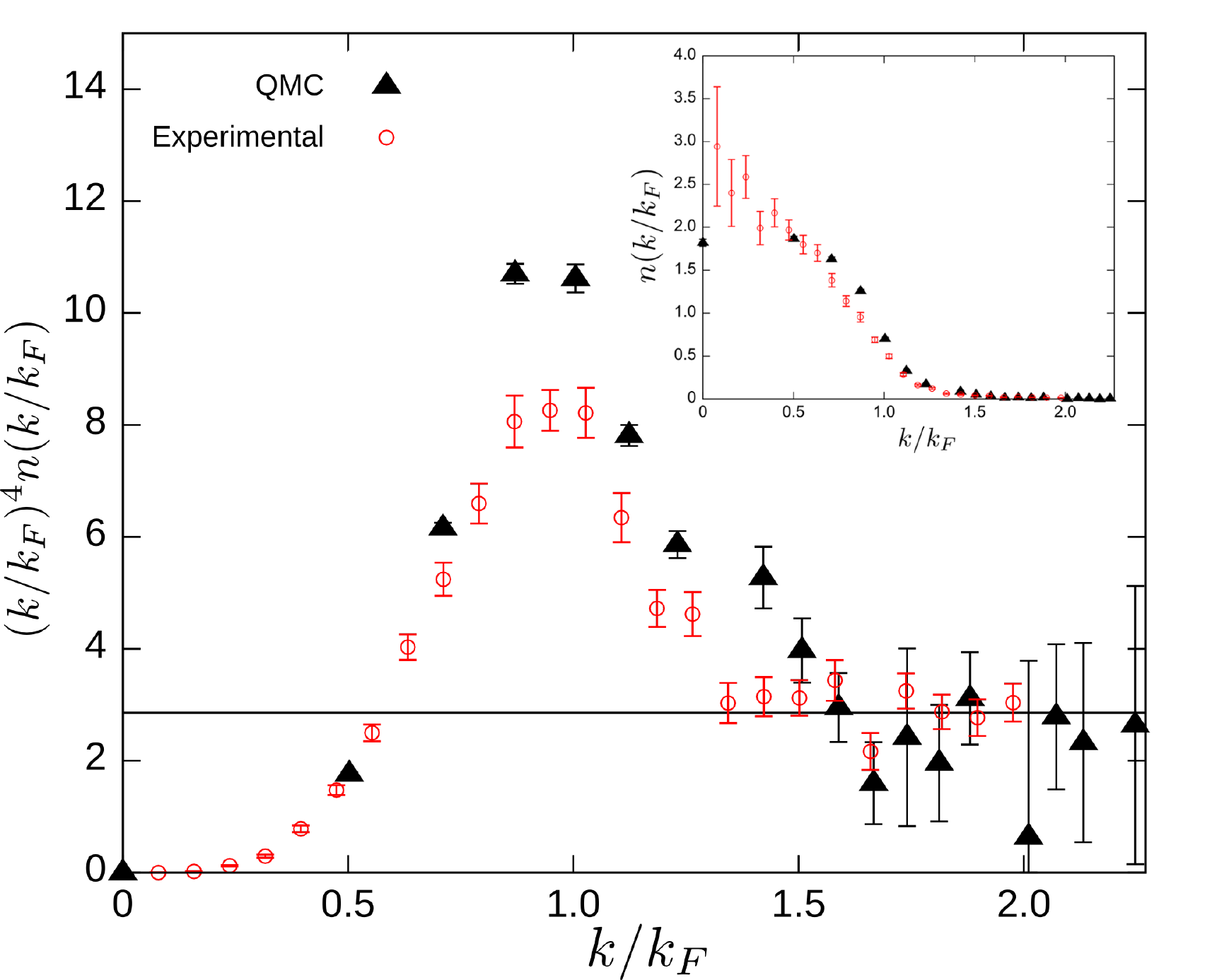}
\caption{\label{Fig:nofk}(Color online)
Momentum density distribution $n(k/k_F)$
multiplied by $(k/k_F)^4$
as a function of $k/k_F$.
In the inset we show  $n(k/k_F)$
as a function of $k/k_F$.
The solid line shows our estimated value of $\mathcal{C}$. The
experimental data is for an inhomogeneous gas \cite{ste10}.
}
\end{center}
\vspace{-5ex}
\end{figure}

The contact $C$ remarkably controls short-distance correlations.  On the
other hand, the momentum distribution $n_{\sigma}(k)$ in the spin state
$\sigma$ for large enough momenta is given by $n_{\sigma}(k) = C / k^4$.
We have computed the quantity $n(k/k_F)(k/k_F)^4$ as a function of
$k/k_F$, and our results are shown in \fig{nofk}.  The contact term is
dominant for momentum states larger than approximately $1.6k_F$, as we can
see from the figure.  This dominance is expected since
$n(k/k_F)(k/k_F)^4 \rightarrow \frac{2}{3\pi^2 Nk_F} \mathcal{C}$.
Although the estimated values of $n(k/k_F)(k/k_F)^4$ are noisy for large
momenta, it is possible to observe a trend towards the value of $\mathcal
C$, estimated from the pair correlation function, and displayed as a solid
line.  The less than optimal agreement of our results with the
experimental data might come from various sources. It might be due to
calculations done at zero temperature while the experiments are, of
course, done at finite temperature. Other possibilities might include the
asymptotic form we have considered for the guide function; it eventually
needs to be improved by including more long-range correlations. However it
is worthwhile mentioning that other DMC calculations~\cite{car03} would
also overestimate the values of $n(k/k_F )$ at low values of $k$.

\section{Summary}

In summary, we have performed for the first time diffusion Monte Carlo
calculations of a system interacting with a contact interaction.
This has allowed us to have a more faithful description of dilute ultracold
Fermi gases at unitarity that opens
possibilities of more accurate and precise calculations of other important
quantities associated with this system. The application of
this approach has allowed us to
compute such quantities as the reduced chemical potential and Tan's contact
parameter in better agreement with experiment than some previous
calculations. We have
introduced an
alternative way of studying ultracold atoms in the unitary limit which
will be of value in the investigation of these systems,
and in other situations where a description using a finite effective
range interaction might be inaccurate.

\begin{acknowledgments}

\textsc{rp} thanks the Department of Physics of the Arizona State University.
\textsc{sav}
is happy to acknowledge the hospitality of \textsc{jila} specifically that of
Ana Maria
Rey and Murray Holland. We thank  Deborah Jin,
Tara Drake and Rabin Paudel
for very useful discussions and also for sharing with us their
experimental values.
This work was partially supported by
the U.S.~Department of Energy, Office of Nuclear Physics, under contract
\smaller{DE-AC52-06NA25396} by the programs \textsc{nuclei
s}ci\textsc{dac} and \textsc{lanl ldrd}, the National Science Foundation
grant \smaller{PHY-1404405} and by grants of the Brazilian agencies
\smaller{FAPESP-2014/20864-2}, \smaller{FAPESP-2010/10072-0};
\smaller{CAPES 11540/13-3} and
\smaller{PVE-087/2012}.
Computational resources have been provided by Los Alamos Open
Supercomputing and \textsc{cenapad-sp} at Unicamp.
We also used resources provided by \textsc{nersc}, which is
supported by the \textsc{us doe} under contract
\smaller{DE-AC02-05CH11231}.
\end{acknowledgments}

\appendix{
\section{Sampling the unitary propagator}
We will begin by looking at the dominant part of the wave function when
an opposite spin pair have a small separation. In this case, we can
approximate the trial wave function ratio as
\begin{equation}
\frac{\Psi_T(R)}{\Psi_T(R')} \sim \frac{r'}{r}
\end{equation}

Since the propagator consists of the free particle gaussian in all
channels except the s-wave, we separate it into the usual
spherical coordinates $r$, $\cos\theta$,
and $\phi$. Starting with the free particle propagator, we take
the $z$ axis along the initial value $\vec r'$ (note we use primed
coordinate for the initial value here for convenience; in the main
text the initial coordinates are unprimed and the sampled coordinates
are primed),
and write the importance sampled gaussian
as
\begin{equation}
\frac{r'}{r}
\frac{1}{(2\pi\sigma^2)^{3/2}} e^{-\frac{|\vec r- \vec r'|^2}{2\sigma^2}}
= 
\frac{r'}{r}
\frac{1}{(2\pi\sigma^2)^{3/2}}
e^{-\frac{r^2+r'^2}{2\sigma^2}}
e^{\frac{rr'}{\sigma^2}\cos\theta} \,.
\end{equation}
Normalizing the $\cos\theta$ part
\begin{equation}
\int_{-1}^1 d\cos\theta e^{\frac{rr'}{\sigma^2}\cos\theta} = 
\frac{2\sigma^2}{rr'}\sinh\left (\frac{rr'}{\sigma^2}\right )
\end{equation}
given an $r$ and $r'$ value, we can sample the angular part from
\begin{eqnarray}
p_\phi(\phi) &=& \frac{1}{2\pi}
\nonumber\\
p_\theta(\cos\theta) &=& \frac{rr'}{2\sigma^2
\sinh\left (\frac{rr'}{\sigma^2}\right )} e^{\frac{rr'}{\sigma^2} \cos\theta}
\,.
\end{eqnarray}
Once we know $r$,
we can sample $\cos\theta$ by sampling a uniform random number $0<\xi<1$
and
\begin{equation}
\cos\theta = 1+\frac{\sigma^2}{rr'}
\ln \left [\xi \left (1-e^{-\frac{2rr'}{\sigma^2}}
\right )
 +e^{-\frac{2rr'}{\sigma^2}} \right ]
\end{equation}

The importance sampled gaussian is now
\begin{equation}
\frac{r'}{r}
\frac{1}{(2\pi\sigma^2)^{3/2}} e^{-\frac{|\vec r- \vec r'|^2}{2\sigma^2}}
= p_\phi(\phi) p_\theta(\cos\theta) 
\frac{1}{\sqrt{2\pi}\sigma r^2}
\left [ e^{-\frac{(r-r')^2}{2\sigma^2}}
-e^{-\frac{(r+r')^2}{2\sigma^2}} \right ] \,.
\end{equation}
The integral of the $r$ part over $r^2 dr$ is
\begin{equation}
\label{eq.normgauss}
\int_0^\infty dr
\frac{1}{\sqrt{2\pi}\sigma}
\left [ e^{-\frac{(r-r')^2}{2\sigma^2}}
-e^{-\frac{(r+r')^2}{2\sigma^2}} \right ]
= {\rm erf}\left (\frac{r'}{\sigma\sqrt{2}}\right ) \,.
\end{equation}

We now look at the ``extra'' piece from the unitary s-wave interaction.
It has the importance sampled form
\begin{eqnarray}
\frac{r'}{r}\frac{\sqrt{2\pi}}{4\pi^2 \sigma rr'}e^{-\frac{(r+r')^2}{2\sigma^2}}
\,.
\end{eqnarray}
Here the angular part is isotropic, so we can sample the angles from
\begin{eqnarray}
p_\phi(\phi) &=& \frac{1}{2\pi}
\nonumber\\
p^{(0)}_\theta(\cos\theta) &=& \frac{1}{2}
\end{eqnarray}
and the importance sampled function becomes
\begin{equation}
\frac{r'}{r}\frac{\sqrt{2\pi}}{4\pi^2 \sigma rr'}e^{-\frac{(r+r')^2}{2\sigma^2}}
=p_\phi(\phi) p^{(0)}_\theta(\cos\theta)
\sqrt{\frac{2}{\pi}}
\frac{1}{\sigma r^2}e^{-\frac{(r+r')^2}{2\sigma^2}}
\end{equation}

The integral of the $r$ part over $r^2 dr$ is
\begin{equation}
\label{eq.normswave}
\int_0^\infty dr
\sqrt{\frac{2}{\pi}}
\frac{1}{\sigma}e^{-\frac{(r+r')^2}{2\sigma^2}}
= 1-{\rm erf}\left (\frac{r'}{\sigma\sqrt{2}}\right ) \,.
\end{equation}
When we add the normalizations of Eqs. \ref{eq.normgauss}
and \ref{eq.normswave}, we get one, since we should get $e^{-E_0\tau}$,
and with the ground-state energy $E_0 = 0$ since there is no bound state
for the unitary gas.

This suggests a way to sample the propagator. We first sample the $r$
value, with probability that the $r$ value was sampled from the gaussian
we sample $\cos \theta$ from $p_\theta(\cos\theta)$, otherwise, we
sample $\cos\theta$ uniformly. In either case, we sample $\phi$ uniformly.

The basic idea below is to sample from a 1-dimensional
gaussian centered at $r'$. This corresponds to the first term of $g_s$.
If the resulting $r$ is greater than zero, then it is a legal value.
With probability given by the ratio of the radial part of $G_0$ divided
by the sampled 1-dimensional gaussian, we take this $r$ as being sampled
from $G_0$, and sample $\cos \theta$ accordingly. If not, the rejected
terms have been sampled from the $e^{-\frac{(r+r')^2}{2\sigma^2}}$,
so they are half of the s-wave term. If the resulting sampled $r$ is negative,
we take its absolute value and it is also sampled from the s-wave term.

Our algorithm is
\begin{itemize}
\item
Sample a $\phi$ uniformly on $0 < \phi < 2\pi$ or equivalent.
\item
Sample a random variate $y$ from a gaussian with mean zero and variance 1.
\item
The $r$ sample is $r=|r'+\sigma y|$.
\item
If $(r'+\sigma y) \le 0$
sample $\cos \theta$ uniformly. The sampling is complete.
\item
If $(r'+\sigma y) > 0$ then sample a random variate $0 < \xi < 1$ uniformly.
\item
If $\xi < e^{-\frac{2rr'}{\sigma^2}}$ sample $\cos \theta$ uniformly, else
sample $\cos\theta$ from $p_\theta(\cos\theta)$.
\end{itemize}

A graph of the various terms is shown in fig. \ref{f1}.
\begin{figure}
\includegraphics[width=.8\textwidth]{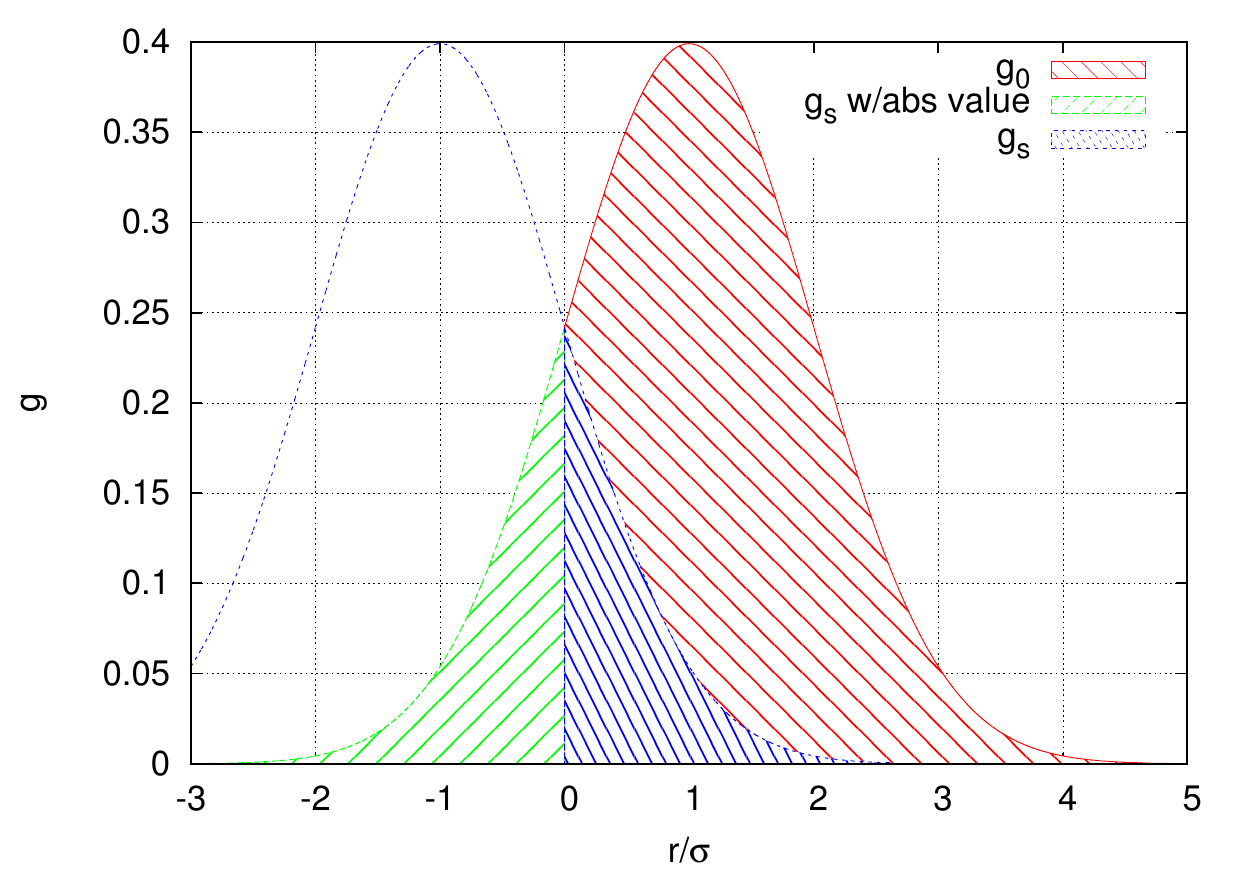}
\caption{
(Color online)
The sampling of the radial part of the propagator is illustrated. We sample
from the gaussian centered around $r'$, in this case taken to be $\sigma$.
The probability of this being from the free gaussian is shown as the
(red) $g_0$ region.
It is the difference between the sampled gaussian and the gaussian
centered around $-r'$, also shown. The (blue) $g_s$ region is half
the probability
of sampling from the s-wave propagator. If we take the absolute value
of the sampled value when it is negative, the samples are from the
(green) $g_s$ w/abs region which gives the other half of the s-wave propagator.
}
\label{f1}
\end{figure}

}


\clearpage

%


\end{document}